\documentclass[journal]{IEEEtran}

\usepackage{tabularx,booktabs}
\usepackage{graphicx}

\usepackage[usenames]{color}
\usepackage{xcolor}

\definecolor{Red}{rgb}{0,0,0}

\ifCLASSINFOpdf
\else
\fi

\begin{document}
\title{Testing optomechanical microwave oscillators for SATCOM application}

\author{Laura~Mercad\'e,
         Eloy~Rico, Jes\'us~Ruiz~Garnica, Juan~Carlos~G\'omez,  Amadeu~Griol, Miguel~A.~Piqueras, Alejandro~Mart\'inez
        and~Vanessa~C.~Duarte%
        %\author{Michael~Shell,~\IEEEmembership{Member,~IEEE,}
        %John~Doe,~\IEEEmembership{Fellow,~OSA,}
        %and~Jane~Doe,~\IEEEmembership{Life~Fellow,~IEEE}%<-this % stops a space
\thanks{L. Mercad\'e, J.C. G\'omez, A. Griol and A. Mart\'inez are with the Nanophotonics Technology Center, Universitat
Politècnica de València, Valencia 46022, Spain and E. Rico, J. Ruiz Garnica, M. A. Piqueras and V. C. Duarte are with DAS Photonics, Valencia 46022, Spain.}
}

\markboth{%Journal of \LaTeX\ Class Files,~Vol.~14, No.~8, August~2015
}%
{Shell \MakeLowercase{\textit{et al.}}: Bare Demo of IEEEtran.cls for IEEE Journals}

\maketitle

\begin{abstract}
The realization of photonic microwave oscillators using optomechanical cavities has recently become a reality. By pumping the cavity with a blue-detuned laser, the so-called phonon lasing regime - in which a mechanical resonance is amplified beyond losses - can be reached and the input signal gets modulated by highly-coherent tones at integer multiples of the mechanical resonance. \textcolor{Red}{Implementing optomechanical cavities on released films with high index of refraction can lead to optical modes at telecom wavelengths and mechanical resonances in the GHz scale, resulting in highly-stable signals in the microwave domain upon photodetection}. Owing to the extreme compactness of such cavities, application in satellite communications (SATCOM) seems highly appropriate, but no experiments have been reported so far. In this paper, an optomechanical microwave oscillator (OMO) built on a micron-scale silicon optomechanical crystal cavity is characterized and tested in a real SATCOM testbed. Using a blue-detuned laser, the OMO is driven into a phonon lasing state where multiple harmonics are generated, reaching tones up to 20 GHz. Under this regime, its practical applicability, remarkably addressing its performance as a photonic local oscillator, has been validated. The results, in addition with the advantages of extreme compactness and silicon-technology compatibility, make OMOs very promising candidates to build \textcolor{Red}{ultra-low} weight photonics-based microwave oscillators for SATCOM applications.

%a novel silicon Optomechanical Oscillator (OMO) is proposed. It is based on a Optomechanical Cavity (OMC) on a silicon chip, in conjunction with a blue-detuned laser.The OMC addresses the interaction between laser light (photons) and mechanical waves (phonons) confined in the cavity, thus creating a frequency comb with armonics spaced by the mechanical frequency (3,874 GHz).  The basic performance of the OMO has been characterized with calculations and experimental setup, obtaining a noise figure as low as -101 dBc/Hz at 100 kHz, which is a remarkable good value for an OptoElectronic Oscillator (OEO) oscillating at GHz frequencies without any feedback mechanism. Its practical applicability has been validated on a real space application test bed. The results, in addition with the advantages of extreme compactness and Silicon-technology compatibility, makes this approach as a very promising candidate to build ultra weight OEOs, highly appropriate for space applications.

% In this paper, we present the practical applicability of a optomechanical microwave oscillator performing around -- bands as a ---- . Furthermore, it has been validated on a real space application test bed. The results, in addition with the advantages of extreme compactness and Silicon-technology compatibility, makes this approach as a very promising candidate to build ultra weight OEOs, highly appropriate for space applications.

\end{abstract}


\IEEEpeerreviewmaketitle

\section{Introduction}

Photonic technology is becoming a key ingredient of the most advanced SATCOM payloads, greatly upgrading their performance thanks to the intrinsic advantages of photonics over electronics: low weight, compactness, low-power consumption, broadband performance and immunity to electromagnetic interference, among \textcolor{Red}{others}. Remarkably, the most important benefits of photonic payload in SATCOM systems have been demonstrated \cite{PIQ19-SPIE,PIQ19-ICSO,POM19-ESTEC} in the framework of European H2020 OPTIMA project \cite{OPTIMA}. As is well known, a local oscillator (LO) is a key building block in microwave -and photonic- satellite payloads based on typical superheterodyne receiver \cite{BEN19-BOOK}. In photonics, this is implemented by an optoelectronic oscillator (OEO) \cite{YAO96-JOSAB} like the SATCOM payload module named photonic frequency generation unit (PFGU) \textcolor{Red}{\cite{BEN19-BOOK}}. Its function is, essentially, to efficiently convert the LO microwave signal into the optical domain.  

\begin{figure}[htbp]
\begin{center}
\includegraphics[width=0.5\textwidth]{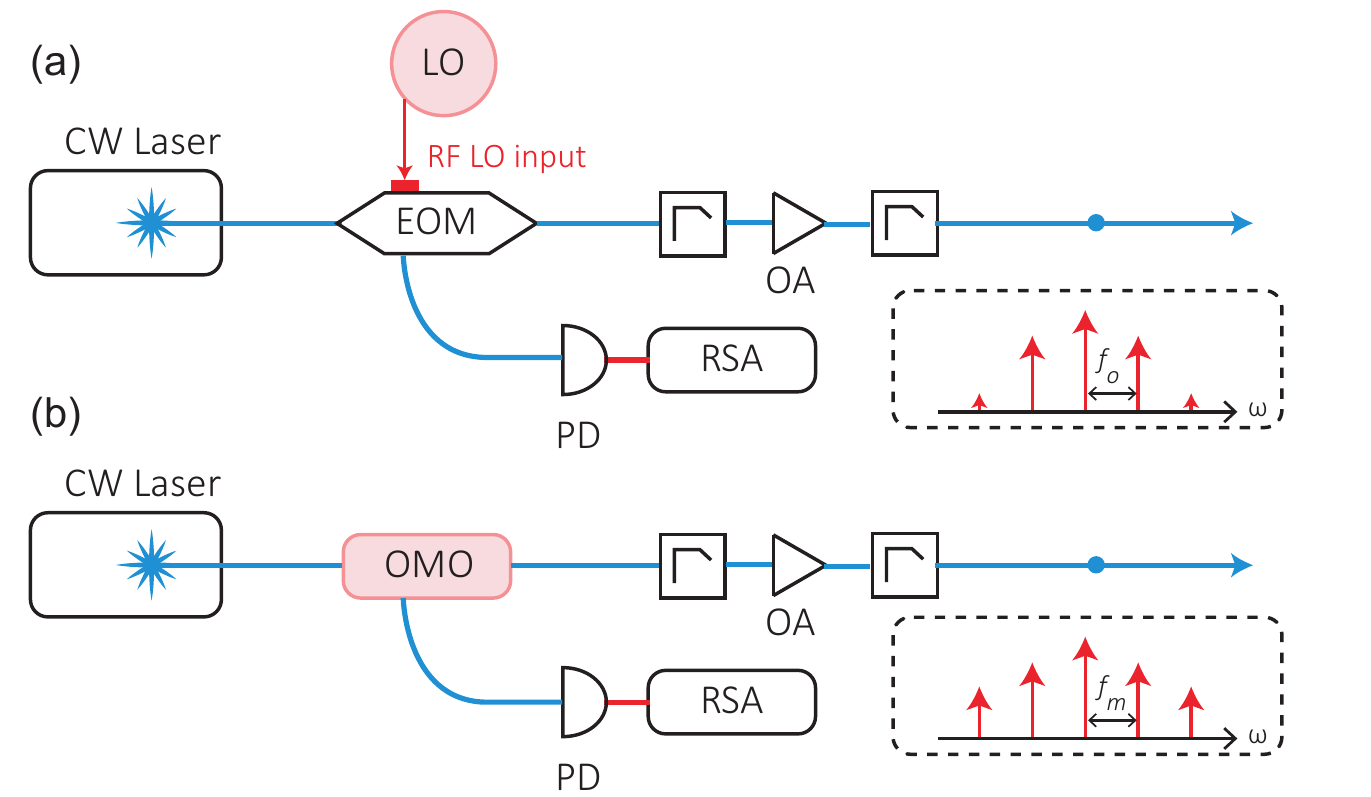}%
\end{center}
\caption{(a) Block diagram of \textcolor{Red}{PFGU} (OEO) of a SATCOM photonic payload. (b) All-optical PFGU approach using an OMO driven to the phonon lasing regime. The used acronyms stands for CW: continuous wave, EOM: electro optical modulator, LO: local oscillator, OA: optical amplifier, PD: photodetector, RSA: RF spectrum analyzer, OMO: optomechanical microwave oscillator.}
\label{fig:1}
\end{figure}

Figure \ref{fig:1}(a) shows the block diagram \textcolor{Red}{and the resulting optical LO signal of a typical PFGU, which includes OPTIMA PFGU}, where an electrical (or RF) LO externally modulates an optical carrier generated by a continuous wave (CW) laser using an electro-optical modulator (EOM), whose output is amplified up to the desired level by an optical amplifier (OA). The main drawbacks of this approach are the need for an electrical oscillator (these LOs are bulky and power consuming) requiring conversion to the optical domain as well as and the need for costly and bulky EOMs, which also require complex control electronics to select and maintain the right operating point.

Alternatively, one may use a fully-optical device as  \textcolor{Red}{PFGU}. The block diagram of this approach, as well as the output optical LO signal, is depicted in Fig. \ref{fig:1}(b). The novel element is an optomechanical cavity \cite{ASP14-RMP}, which under the appropriate conditions can perform as a photonic LO \cite{MER20-NN} within a foot-print of several square microns, thus being suitable for massive integration in silicon wafers. In this kind of cavity, optical and mechanical waves can be simultaneously localized and, as a result of different surface and volume acousto-optic effects \cite{PEN14-NP}, interact strongly, which gives rise to intriguing phenomena. Indeed, upon blue-detuned laser driving, dynamical back-action in the cavity \cite{KIP08-SCI} can lead to self-sustained oscillations, also known as phonon lasing \cite{NAV14-AIPA}. In this regime, the input laser gets modulated by a very narrow tone so the device can perform as a photonic microwave oscillator with frequencies determined by the mechanical resonance(s) of the cavity \cite{HOS10-IEEEJSTQE,GHOR19-ARX,MER20-NN}. Under stronger pumping, the input laser signal gets modulated by a set of narrow peaks at integer multiples of the mechanical resonance frequency, giving rise to the optomechanical (OM) equivalent of an optical frequency comb (OFC) \cite{MER20-NN}. 

%This OM crystal cavity design has been already employed as a photonic frequency conversion of orthogonal frequency division multiplexing microwave signals \cite{MER21-LPR}, thus exhibiting its potential to be a key part of aircraft and satellite communication systems. 
%Owing to its intrinsic nonlinear nature, the cavity also performs as a mixer, as recently demonstrated \cite{MER21-LPR} thus  exhibiting  its  potential  to  be  a  key building block of aircraft and satellite communication systems with photonic payload.

However, despite its appealing features as photonic microwave oscillator \textcolor{Red}{\cite{LIU14-JLT,HUA20-JLT}}, OMO devices have not yet been tested when forming part of complex system used in real applications. In this work, we validate the performance of \textcolor{Red}{a silicon-chip OMO having a fundamental mechanical oscillation frequency around 4 GHz as a photonic LO in a real SATCOM test bed, thus unveiling the potential to use OM devices out of the lab}. 

\section{OM cavity characterization}

\begin{figure}[htbp]
\begin{center}
\includegraphics[width=0.5\textwidth]{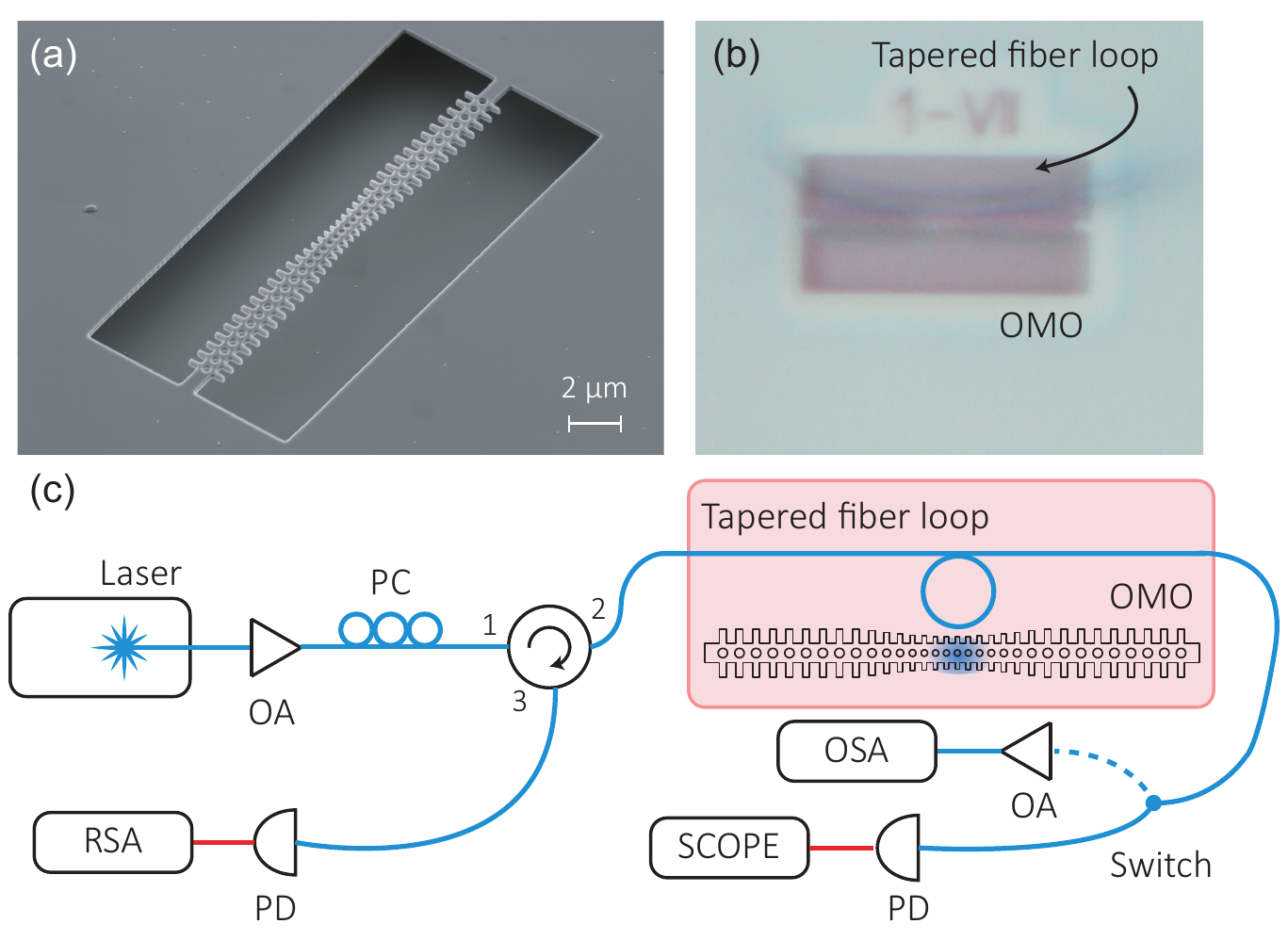}%
\end{center}
\caption{(a) Scanning electron microscopy image of a single OM crystal cavity as the one used in this work. (b) Optical microscope top-view of the coupling between a tapered fiber loop close to the OM cavity. (c) Block diagram of the experimental setup used in the OM characterization of the cavities.}
\label{fig:2}
\end{figure}

Here we describe the silicon OM cavity that we use in our experiments (see Fig. \ref{fig:2}(a)). As reported in \cite{MER20-NN}, the OM cavity essentially consists of a released silicon nanobeam that is structured so that at its sides there are periodic photonic crystal mirrors that provide a photonic bandgap for TE-like modes at the desired wavelengths. The unit-cell dimensions are adiabatically modified when moving from the cavity sides to its center - for instance, the width of the lateral wings is diminished - resulting in localized photonic states with large optical Q factor. This will also result in localized mechanical states \cite{EIC09-NAT}, which can interact with the confined optical modes. 

%These are the mirrors that enable light confinement and whose dimensions are in the order of 20 $\mu$m of length (see caption of Fig. \ref{fig:2}). Also, the dimensions of the mirror unit cell are adiabatically modified when moving towards the centre of the cavity to ensure a large Q factor.   %In the centre of the cavity the dimensions are defined as seen in the caption of Fig. 3 for the defect unit cell, whilst there is adiabatic transition between the mirrors and the centre.

An important point is that we cannot choose the mechanical frequencies \textit{at will}. Instead, they are constrained by our choice of photonic cavity which must resonate at 1550 nm in order to use telecom optical equipment for characterization. Since the light speed is about five orders of magnitude higher than the speed of sound in silicon, mechanical resonances will occur at frequencies five orders of magnitude lower than the optical frequencies \cite{PEN14-NP}. Therefore, OM cavities in silicon nanobeams have mechanical modes with frequencies in the scale of several GHz. Moreover, usually a lot of mechanical modes, with different localization patterns will coexist \cite{GOMIS14-NCOMM} (notice that extended modes with frequencies of tens of MHz will also be present \cite{NAV15-SR}). In our cavity, we get a set of GHz-scale mechanical modes confined in the defect region \cite{MER21-PRL}. However, according to experiments, the mode with the largest OM interaction - see below - is the one which oscillates at a mechanical frequency \textcolor{Red}{$f_{m}=\Omega_{m}/2\pi$} =3.874 GHz \cite{MER20-NN}. \textcolor{Red}{Other OM cavities in silicon nanobeams can lead to more mechanical frequencies in the GHz regime, even reaching frequencies above 10 GHz through more sophisticated designs \cite{REN20-NCOMM}}.

From the optical and mechanical field patterns, which can be obtained via numerical simulations \cite{OUD14-PRB}, it is possible to calculate the vacuum OM coupling rate, \textcolor{Red}{$g_{0}=g_{OM}\sqrt{\hbar/2m_{eff}\Omega_{m}}$}, which \textcolor{Red}{measures the shift of the optical frequency due to the fluctuations of the ground state of the mechanical oscillator \cite{ASP14-RMP, PEN14-NP}. In the last definition, $\hbar$ is the reduced Planck constant and $m_{eff}$ is the effective mass of the cavity. Such coupling rate, $g_{OM}$, has two main contributions, the moving interface (MI) and photoelastic (PE) effect, which model respectively surface and volumetric OM interaction, resulting in $g_{OM}=g_{MI}+g_{PE}$. The MI term can be calculated as \cite{CHAN12-PHD}:}

\begin{equation}
\textcolor{Red}{g_{MI}=-\frac{\omega}{2}\frac{\oint (\textbf{U}\cdot\textbf{\^n})(\Delta\varepsilon\vert\textbf{E}_{\parallel}\vert^{2}-\Delta\varepsilon^{-1}\vert\textbf{D}_{\perp}\vert^{2} dS)}{\int \textbf{E}\cdot\textbf{D} dV}}
\end{equation}

\noindent \textcolor{Red}{where $\textbf{U}$ is the normalized displacement field (max{$\vert\textbf{U}\vert$}=1), $\textbf{\^n}$ is the outward pointing normal to the boundary, $\Delta\varepsilon=\varepsilon_{Si}-\varepsilon_{Air}$ and $\Delta\varepsilon^{-1}=\varepsilon_{Si}^{-1}-\varepsilon_{Air}^{-1}$ \cite{OUD14-PRB}. Here, $\varepsilon_{Si}$ and $\varepsilon_{Air}$ accounts for the silicon and air permittivity, respectively.}

\textcolor{Red}{The PE coupling rate can be calculated as:}

\begin{equation}
\textcolor{Red}{g_{PE}=-\frac{\omega}{2}\frac{\langle E\vert \delta\varepsilon\vert E\rangle}{\int \textbf{E}\cdot\textbf{D} dV}}
\end{equation}

\noindent \textcolor{Red}{with $\delta\varepsilon=-\varepsilon_{0}n^{4}p_{ijkl}S_{kl}$. Here, $\varepsilon_{0}$ is the vacuum permittivity, $n$ the refractive index of silicon, $p_{ijkl}$ are the photoelastic tensor components and $S_{kl}$ the strain tensor components.}

The vacuum OM coupling rate is one of the key parameters of an OM cavity: the higher it is, the more efficient all the underlying OM phenomena will be. In our case, we get $g_{0}/2\pi$ = 540 kHz  \cite{MER20-NN}, not far from the best values reported in literature \cite{REN20-NCOMM,CHAN12-APL}.

%Other parameters can also be calculated (see details in \cite{MER20-NN}) but many of the most relevant input required for simulations at the system level (optical power modulation index, RF power) can only be obtained from experiments since they depend strongly on the experimental conditions (namely, the input laser power or the positioning of the excitation fibre loop with respect of the OM cavity). 

Essentially, under the appropriate conditions (blue-detuned driving to achieve phonon lasing), at the output of the OM cavity we will have an optical carrier (provided by the driving laser and with frequency $\omega_{L}$) modulated by a set of harmonics spaced by the mechanical frequency. Each one of these peaks will have a certain amplitude or power, which will determine the available microwave power $P_{LO,n}$ at a certain frequency $f_{LO,n}=n\omega_{m}/2\pi$. As shown below, these microwave tones will display very low \textcolor{Red}{phase noise}, in particular the 1st harmonic \cite{MER20-NN}.

Once light is coupled into the cavity via a tapered fiber loop (see Fig. \ref{fig:2}(b)), OM interaction will transduce the mechanical modes into the optical signal, being then detected using a photodetector. The full experimental setup to measure the OM cavity is presented in Fig. \ref{fig:2}(c) (the loop and the cavity are not to scale). Here, a continuous wave (CW) is generated with a tuneable fiber-coupled external cavity diode laser (New Focus Velocity TLB-6728). Then, the optical signal is amplified by means of an erbium doped fiber amplifier (EDFA) to fulfil our power requirements, and a polarization controller (PC) sets the required polarization that will finally arrive at the device under characterization. \textcolor{Red}{Driving the cavity requires to have the incident field parallel to the chip plane so that a transverse electric (TE) mode of the cavity is excited \cite{MER20-NN}.} After the polarization controller the optical wave is sent through a circulator that generates two paths: transmission (port 2) and reflection (port 3). To ensure a closed environment that minimizes external interferences, the whole system is placed inside a methacrylate cage structure where both the tapered micro-looped fiber and the sample are located. 

When the laser is switched on, the light travels along the fiber and couples to the cavity -via evanescent field coupling- provided that the loop is close enough to it. The relative position between these two elements can be accurately tuned with piezoelectric controllers placed at the sample holder where the OM cavity is located. Then the cavity can scatter lights back into the fiber, being guided along both the transmission and reflection direction. Regarding the transmission channel, we can switch between two different configurations as a function of the measurement to be performed. For optical spectrum measurements, the signal is first amplified (if needed) with an EDFA and then it is analyzed through an optical spectrum analyzer (OSA) ANDO AQ6317C. The optical resonance measurements are recorded using an alternative configuration. In this case, the signal is attenuated with a variable optical attenuator (VOA), and then it is photo detected via a switchable gain amplified detector (Thorlabs PDA20CS-EC - InGaAs switchable gain amplified detector) and finally recorded with a scope (DSO81204B Infinium high performance oscilloscope). 

The reflection channel ends in either a 12 GHz-bandwidth photodetector (New Focus 1544-B DC-coupled NIR fiber-optic receiver) or a 50 GHz-bandwidth photodetector (u2XPDV2120R). It must be noted that whilst the 12 GHz bandwidth detector has more sensitivity, the second one permits us to measure higher-order harmonics of the fundamental mechanical resonance. Once the signal has been photodetected we can choose either to analyze it through an RF spectrum analyzer (RSA) or use it as the trigger signal for the temporal trace measurements. It must be remarked that different RSA have been employed in our experiments: one for the analysis of the mechanical response of the cavity (Rohde $\&$ Schwarz FSQ 40) and another for the phase noise measurements (Rohde $\&$ Schwarz FSUP signal source analyzer or Anritsu MS2850A signal analyzer). 

%As presented above, a close view of the coupling scheme between the optomechanical crystal cavity (OMC) and the tapered fiber loop is shown in Fig. \ref{fig:2}(b). The relative position between the sample and the loop can be tuned through the stage where the sample holder is placed, where a fine tune can be performed via the above-mentioned piezoelectric controllers. 

\section{Optomechanical microwave oscillator (OMO)}

\begin{figure}[htbp]
\begin{center}
\includegraphics[width=0.5\textwidth]{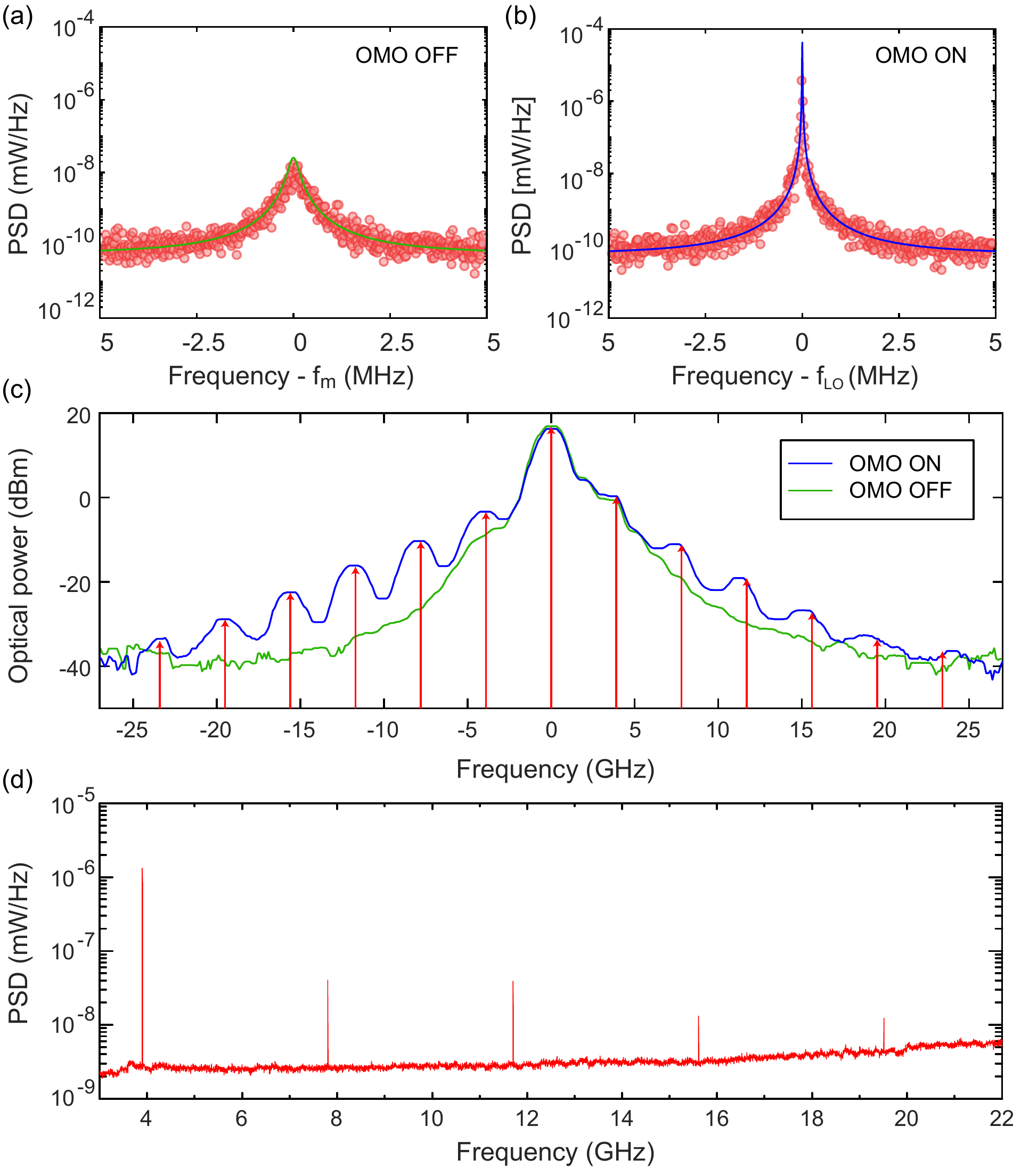}%
\end{center}
\caption{(a) Power spectral density of the transduced mechanical spectrum when the \textcolor{Red}{OMO is off (OMO OFF) and the tranducd signal comes from the thermal occupation. (b) PSD spectrum of the first harmonic when the \textcolor{Red}{phonon lasing state} is reached (OMO ON). (c) Recorded optical spectrum when the OMO is activated or not (in green and blue, respectively) showing a set of peaks corresponding to different harmonics represented (in red) at the expected position. (d) PSD spectrum of the OFC generated with the OMO.}}
\label{fig:3}
\end{figure}

When the cavity is driven by a low-power laser whose wavelength is blue-detuned with respect to the cavity resonance, the thermal motion of the cavity is transduced into the optical signal and can be detected in the photodetectors. An example of the transduced mechanical motion response before reaching the instability threshold is presented in Fig. \ref{fig:3}(a), which represents the thermal phonons present in the cavity and moving at frequencies around the mechanical frequency $f_{m}$, which in this particular device was $f_{m}\approx$ 3.894 GHz. The spectrum in Fig. \ref{fig:3}(a) has been labelled as OMO OFF because the mechanical mode is simply transduced but has not yet reached the self-oscillations regime. When driving the cavity with a blue-detuned laser with enough power, the overall mechanical damping rate is highly reduced. By these means, the mechanical cavity dynamics evolves and amplification of the mechanical motion can be obtained as a result of dynamical back-action \cite{NAV14-AIPA}. Indeed, a threshold of instability can be attained once the mechanical losses equate to the optical \textcolor{Red}{antidamping} and the oscillator reaches a condition of self-oscillation \textcolor{Red}{(also known as phonon lasing regime)}. \textcolor{Red}{That} results in a very narrow tone in the detected spectrum at the mechanical resonance as shown on Fig. \ref{fig:3}(b) (labelled as OMO ON), which depicts the electrical spectrum around the central frequency of the detected peak (termed $f_{LO}$ hereinafter) once we have reached the phonon lasing condition. The narrowing of the detected signal, as well as the increase of the power, both signatures of phonon lasing, can be easily noticed. This is a direct proof that the cavity can be employed as an OMO. It must be highlighted that despite its equivalence with the OEO in terms of the functions it can provide, here, the involved mechanism is a self-sustained oscillation originated from OM interaction.

 Notice that when back-action effects take place, the detected frequency in the electrical spectrum can be different from the original mechanical frequency. This shift is a result of the optical spring shift effect \cite{REN20-NCOMM, NAV14-AIPA}, which changes the stiffness of the silicon nanobeam according to the total amount of power in the cavity \textcolor{Red}{and the laser-cavity detuning}. This means that in the phonon lasing regime $f_{LO}$ will slightly differ from $f_{m}$, which in real applications can be used for dynamic tuning of the oscillator. \textcolor{Red}{In the cavity under study, the variation range of $f_{LO}$ can be up to 5 MHz once the cavity is in the lasing regime \cite{MER20-NN,MER21-LPR}.}

%as well as integer multiples of it. This narrow tone can, therefore, act as an embedded LO oscillating at integer multiples of the fundamental mechanical resonance, amongst other applications. A broad description of this phenomenon - termed optomechanical frequency comb - can be found in \cite{MER20-NN}.

Indeed, under sufficiently strong pumping \cite{MER20-NN}, the optical carrier becomes modulated by a set of harmonics spaced \textcolor{Red}{by} $f_{LO}$ thus generating an \textcolor{Red}{optical frequency comb of OM nature}. At this operation point, labelled as OMO ON, the resulting optical spectrum of the transmitted signal is depicted in Fig. \ref{fig:3}(c). Here, a noticeable difference between the optical response when the OMO is activated (blue curve) or not (green curve) can be appreciated. In the case when the mechanical motion is only transduced - OMO OFF case - only two sideband peaks spaced $\pm f_{m}$ from the CW laser signal are observed. Once the cavity is in the self-sustained regime, different sidebands (depicted in red arrows) at frequencies $n\times f_{LO}$ around the laser frequency, where $n$ is the harmonic index, are generated. If we analyze the RF spectrum or power spectral density (PSD) of the system when the OMO is active, obtained upon photodetection of the previously analyzed optical signal, higher-order harmonics appear, as presented in Fig. \ref{fig:3}(d) for the first five harmonics.

A key aspect determining the quality of microwave oscillators is the presence of random fluctuations in the phase of the generated tone. In the frequency-domain representation, these fluctuations can be measured by the phase noise, which is related to the time domain deviations from the case of an ideal signal with zero jitter. In our case, once the cavity is in the lasing regime, the generated microwave tone can be characterized by these means. \textcolor{Red}{The phase noise of the different harmonics generated by driving the cavity in the lasing state was reported in \cite{MER20-NN}, showing a behavior similar to that of standard harmonic mixers (the phase noise grows with the order of the harmonic). However, whilst the phase noise is a powerful tool to analyze the short-term stability of the oscillator, the characterization of its long-term stability is also needed for practical applications. Such long-term stability is usually quantified by the Allan deviation $\sigma_{y}(\tau)$, which provides information about the frequency drift of the oscillator \cite{RUB08-BOOK, LUAN14-SR, YU16-NCOMM}}. 

%In such a way, we measured the phase noise of the generated microwave tone as well as several harmonics ($n\times f_{LO}$) at a frequency offset of 100 kHz. In principle, the harmonic mixing process should result in an added phase noise of $20\times log(n)$ with respect to that of the first harmonic. As shown in Fig. 7, the previous rule is well satisfied in our device. Thus, the OM cavity acts as a harmonic generator (comb) and the phase noise behaves as in a conventional harmonic mixer, being the phase noise impaired by the factor previously mentioned. Even though the harmonic mixing process results in an added phase noise of $20\times log(n)$ experimental measurements show noise figures $<$ -80 dBc/Hz above 15 GHz.

%esides the OMO short-term stability analysis given through the phase noise analysis, a long-term stability characterization may also be needed for practical applications. The long-term stability can be quantified by the Allan deviation $\sigma_{y}(\tau)$ term for long averaging times, which can provide information about the frequency drift of our oscillator. To that end, $\sigma_{y}(\tau)$ can be defined as the variance of the difference between two fractional frequency values measured at consecutive times. 

\begin{figure}[htbp]
\begin{center}
\includegraphics[width=0.5\textwidth]{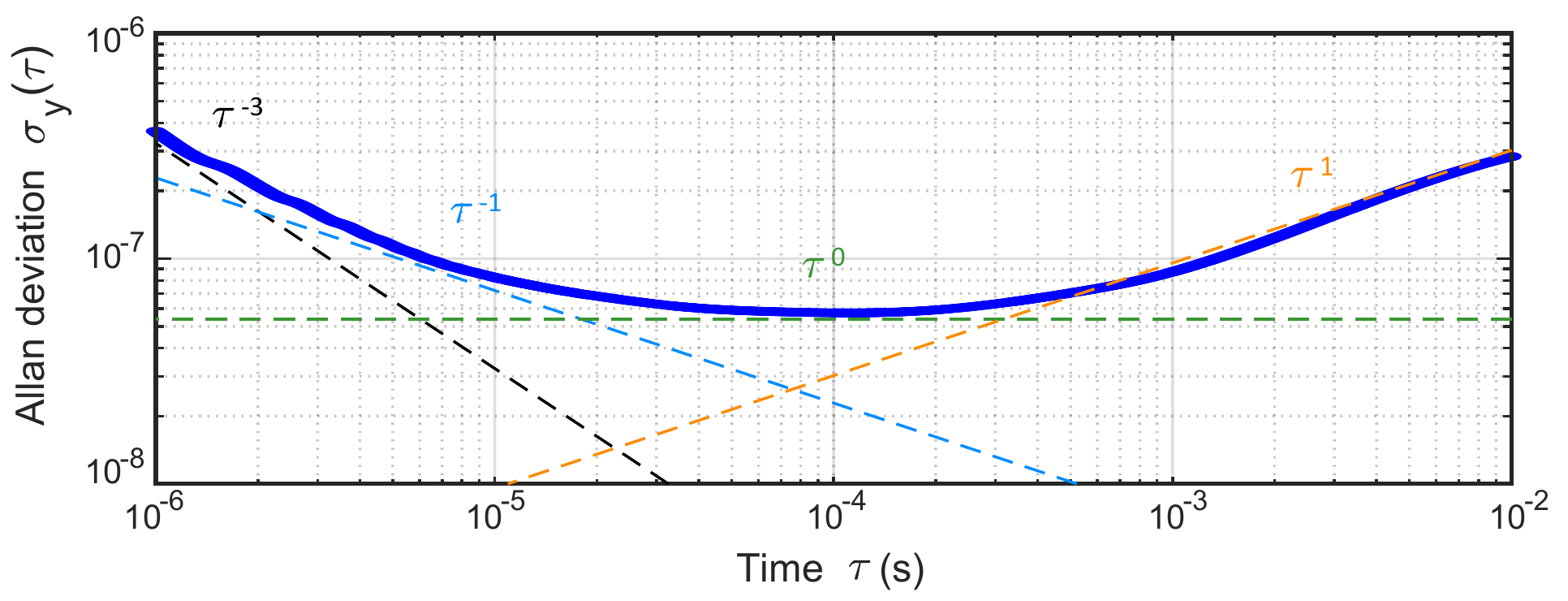}%
\end{center}
\caption{Allan deviation results converted from the measured phase noise results for the OMO under study (blue continuous line) and its power law fits contributions (dashed lines).}
\label{fig:4}
\end{figure}

The Allan variance $\sigma_{y}^{2}(\tau)$ depends on the variable $\tau$ and expresses the mean square of all frequency counter samples separated in time by $\tau$ over the entire measurement interval \cite{RUB08-BOOK}. The resulting $\sigma_{y}(\tau)$ is presented at Fig. \ref{fig:4}. Here, different noise sources as the flicker frequency noise ($\sigma_{y}(\tau)\approx\tau^{0}$) and a contribution of the random walk of frequency ($\sigma_{y}(\tau)\approx\tau^{1}$) can be appreciated. \textcolor{Red}{The Allan deviation has been calculated through:}

\begin{equation}
\sigma_{y}(\tau)=\sqrt{\int_{0}^{\infty}\frac{4f^{2}L(f)}{f_{c}^{2}} \frac{\sin^{4}(\pi\tau f)}{(\pi\tau f)^{2}}df}
\end{equation}

\textcolor{Red}{where $L(f)$ is the measured phase noise, $\tau$ the averaging time and $f_{c}$ the carrier frequency of the oscillator under study \cite{MER21-PRL,LUAN14-SR}. The derived Allan deviation through the phase noise data is in agreement with the power-law fits (in dashed lines in Fig. \ref{fig:4} obtained via the conversion relationships in \cite{RUB08-BOOK}, thus resulting in a good approximation.} \textcolor{Red}{On the other side,} a broader analysis of the long-term stability of this system is presented in Ref. \cite{MER21-LPR}.

%por aqui

\begin{figure}[htbp]
\begin{center}
\includegraphics[width=0.5\textwidth]{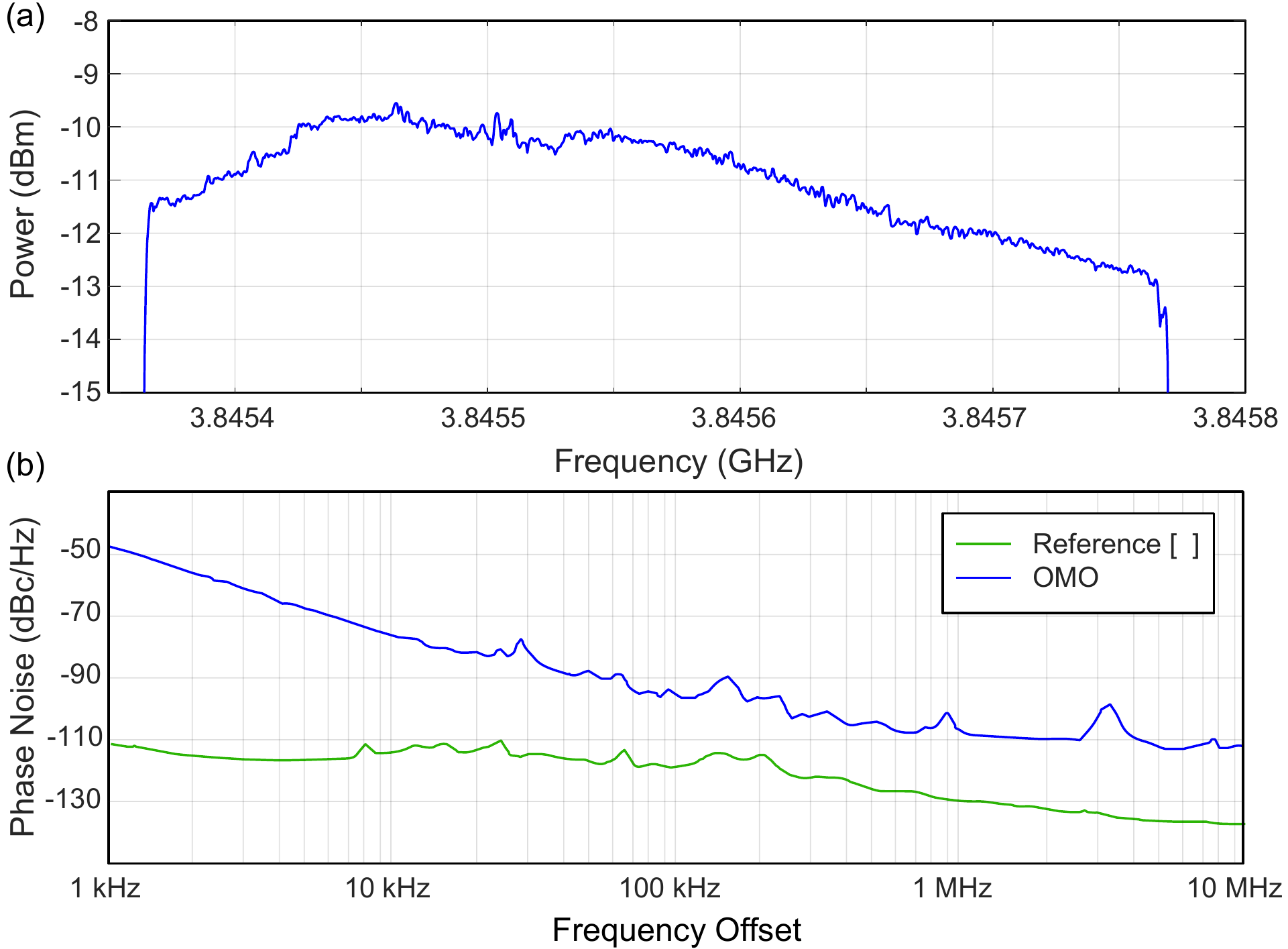}%
\end{center}
\caption{(a) \textcolor{Red}{Maximum held trace showing the amplitude} stability and temporal frequency drift of the 1st harmonic during a 3 hours measurement. (b) Phase noise (PN) level of OMO (blue trace) versus a commercial low PN microwave generator (green trace).}
\label{fig:5}
\end{figure}

To further assess the stability of the detected electrical signal for its use as an LO, the long-term signal stability was measured. This measurement was carried out by letting on trace in max-hold mode, so it maintains the maximum power value for the rest of the test. Figure \ref{fig:5}(a) shows the power detected for first harmonic over a 3 hours time window measurement. Also, the single-sideband (SSB) phase noise was obtained (blue curve) using the phase noise measurement application of the spectrum analyzer. As a reference, an ultra-low phase noise RF generator was employed (green curve). Figure \ref{fig:5}(b) shows both responses. \textcolor{Red}{Two different OMO configurations were measured to discard phase noise high dependence on each one. It should be considered that the OMO is under different configuration since it was measured under different conditions, such as: the wavelength detuning or input power.} The phase noise level at a 100 kHz offset frequency shown in Fig. \ref{fig:5}(b) is consistent with the phase noise obtained in \cite{MER20-NN} for a similar device.

\section{OMO integration in a SATCOM testbed}

\textcolor{Red}{Once the OMO has been validated and its basic performance has been characterized, it is time to assess its practical applicability on a real space-application testbed, like in the }context of the OPTIMA project \cite{OPTIMA}, \textcolor{Red}{where the main goal was demonstrating the strengths of photonic payloads, as reported before. The architecture of photonic satellite payload includes LO generation (with the requirements of high power and low phase noise \cite{KAR19-IEEE}),  frequency down conversion (DOCON) and photonic receiver, with intermediate frequency (IF) as an output [1]. The reference specifications of OPTIMA project are summarized in Table \ref{tab:1}.} In our case, the goal is to test the performance of the OMO device presented in Ref. \cite{MER20-NN} within the OPTIMA system.

\begin{table}[h]
  \centering
   \caption{OPTIMA test bed reference specification.}
   \label{tab:1}
  \begin{tabular}{@{}cccccc@{}}
    \toprule
Parameter & Requirement \\ 
    \midrule
 Gain (dB) & $<$ -17 \\ 
 Noise Figure (dB) & $<$ 42 \\
 Phase Noise (dBc/Hz) &  \\
 10 Hz &  $<$ -39\\
   100 Hz &  $<$ -73\\
   1 kHz &  $<$ -85\\
   10 kHz &  $<$ -96\\
   100 kHz &  $<$ -105\\
   1 MHz &  $<$ -114\\
   10 MHz &  $<$ -131\\
 LO Harmonics (dBc) & $<$ -40 max \\
 Intermodulation products (dBc) & $<$ -65 max \\
    \bottomrule
  \end{tabular}
 
\end{table}

 \begin{figure*}[t]
%\begin{center}
\includegraphics[width=\textwidth]{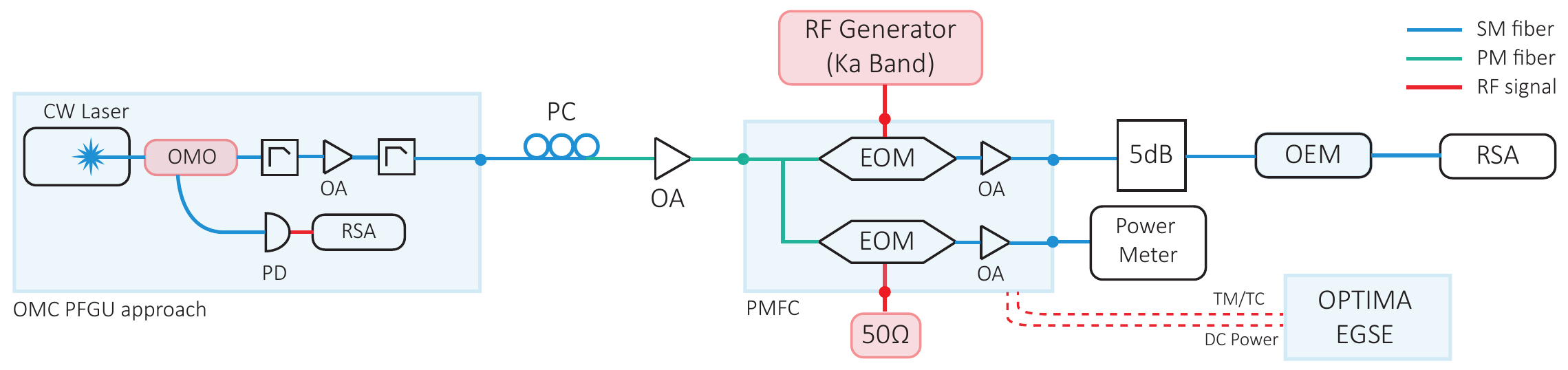}%
%\end{center}
\caption{\textcolor{Red}{Integration of the OMO device in the OPTIMA testbed. The used acronyms stands for CW: continuous wave, OMO: optomechanical microwave oscillator, OA: optical amplifier, PD: photodetector, RSA: RF spectrum analyzer, PC: polarization controller, EOM: electro optical modulator, OEM: opto electronic module, OMC: opto mechanical cavity, PFGU: photonics frequency generation unit, PMFC: photonic multi-frequency converter, TM/TC: telemetry / telecommand, SM: single mode, PM: polarization-maintaining and EGSE: electrical ground support equipment.}}
\label{fig:6}
\end{figure*}

The testbed is schematically shown in Fig. \ref{fig:6}, where the OMO is used to generate LO signal in an all-optical way. A polarization controller \textcolor{Red}{(PC)} is introduced between the OMO and the \textcolor{Red}{DOCON named} PMFC (photonic multi-frequency converter) unit to properly align the correct polarization axis to the Mach-Zehnder EOM inside the PMFC. \textcolor{Red}{Notice} that the OMO uses a single mode fiber but polarization-maintaining fiber is needed at the input of EOM. Then, an optical amplifier overcomes the power limitation of the OMO (there is a maximum power that we can inject into the OM cavity without damaging the material). Next, a 5dB optical attenuator is placed between the PMFC output fiber and the \textcolor{Red}{the receiver named} OEM \textcolor{Red}{(optoelectronic module)} to prevent optical overpower. Also, the second channel of the PMFC is connected \textcolor{Red}{as a power monitor}. For the sake of simplicity, since only one wavelength is employed, both \textcolor{Red}{multiplexing/demultiplexing} unit and the switching matrix are not mounted on the validation testbed.

\begin{figure}[t]
%\begin{center}
\includegraphics[width=0.5\textwidth]{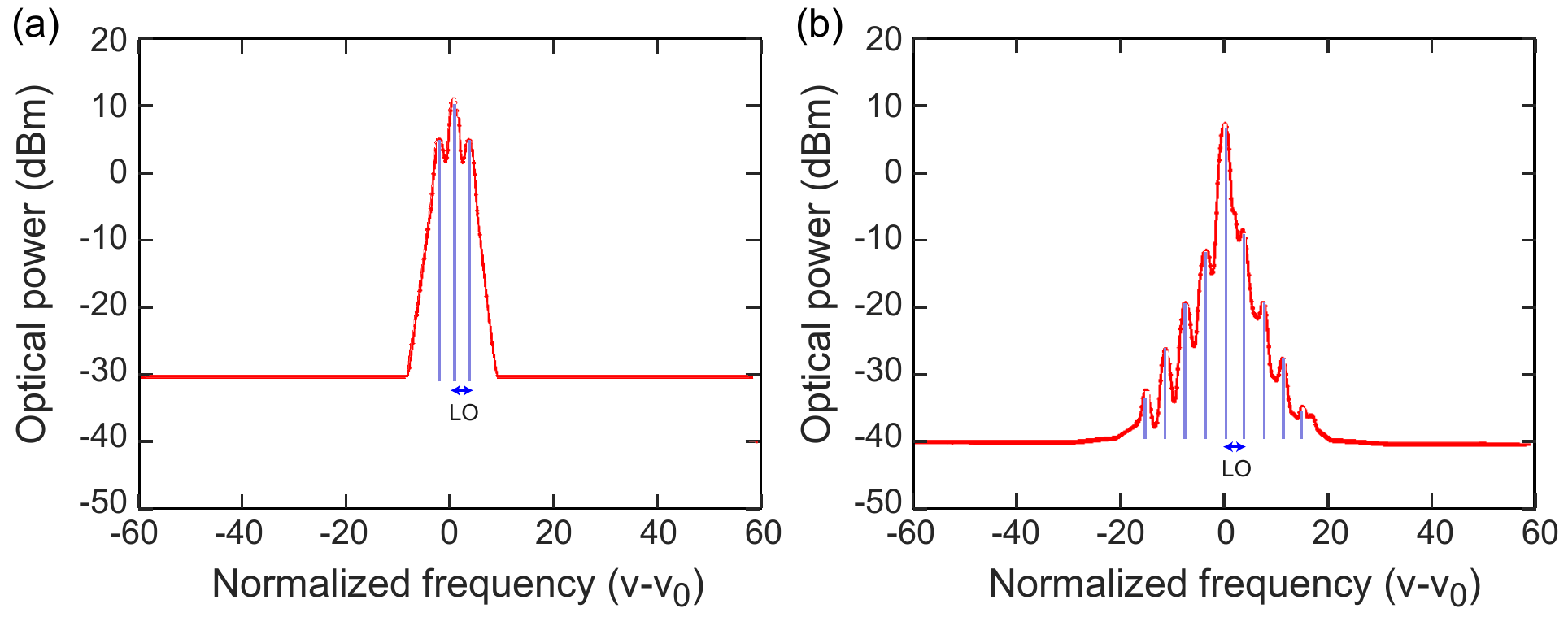}%
%\end{center}
\caption{(a) Optical spectrum \textcolor{Red}{(in red)} at OPTIMA PhLO output. (b) Optical spectrum \textcolor{Red}{(in red)} at OMO output. \textcolor{Red}{Vertical lines represent the different harmonics.}}
\label{fig:7}
\end{figure}

Before showing the functional results, it is important to know the limitations of the OMO in terms of power availability and modulation depth. Figure \ref{fig:7} shows a comparison of the output optical field of both the OMO experiment and the OPTIMA PFGU (also named photonic local oscillator, PhLO, \textcolor{Red}{Fig. \ref{fig:1}(a))}. Specifically, Fig. \ref{fig:7}(a) shows the optical spectrum of the photonic LO from the SATCOM system whereas Fig. \ref{fig:7}(b) depicts the optical spectrum from the OMO when operated to generate an OFC.

%\begin{figure*}[t]
%\center
%\includegraphics[width=\textwidth]{Figure1-2.pdf}
%\caption{prueba}
%\label{fig:intro}
%\end{figure*}

After analyzing such figures, three important aspects must be highlighted. The first is \textcolor{Red}{that} the OMO delivers less total optical power than the OPTIMA PhLO since the carrier and the first-order sideband give +8 dBm and -8 dBm, respectively. On the other hand, the PhLO gives +10 dBm on the carrier and +4.5 dBm on the first-order sideband. Thus, maintaining the rest of the parameters equal for both configurations, higher conversion gain is expected from the OPTIMA PhLO than in the OMO configuration. The second is that the OPTIMA PhLO exhibits a higher modulation efficiency than the OMO, given the fact that the first-order sidebands have more power density than the same sidebands in the OMO. This also contributes to a higher conversion gain in the case of the PhLO. The third observation is that the OMO is more non-linear than the PhLO. This is because the PhLO uses an optical modulator with a medium modulation depth and its behaviour is quite linear, providing most of the modulating power to the fundamental harmonic signal (first sideband). On the contrary, the OMO has a more non-linear response, generating more sidebands so that the power is more distributed among all the harmonic levels. \textcolor{Red}{This means that the OMO will beat the OPTIMA PhLO architecture (keeping the rest of the parameters equal in both architectures) when addressing frequency conversion beyond the first harmonic}.

\section{Test results}

\begin{figure}[t]
%\begin{center}
\includegraphics[width=0.5\textwidth]{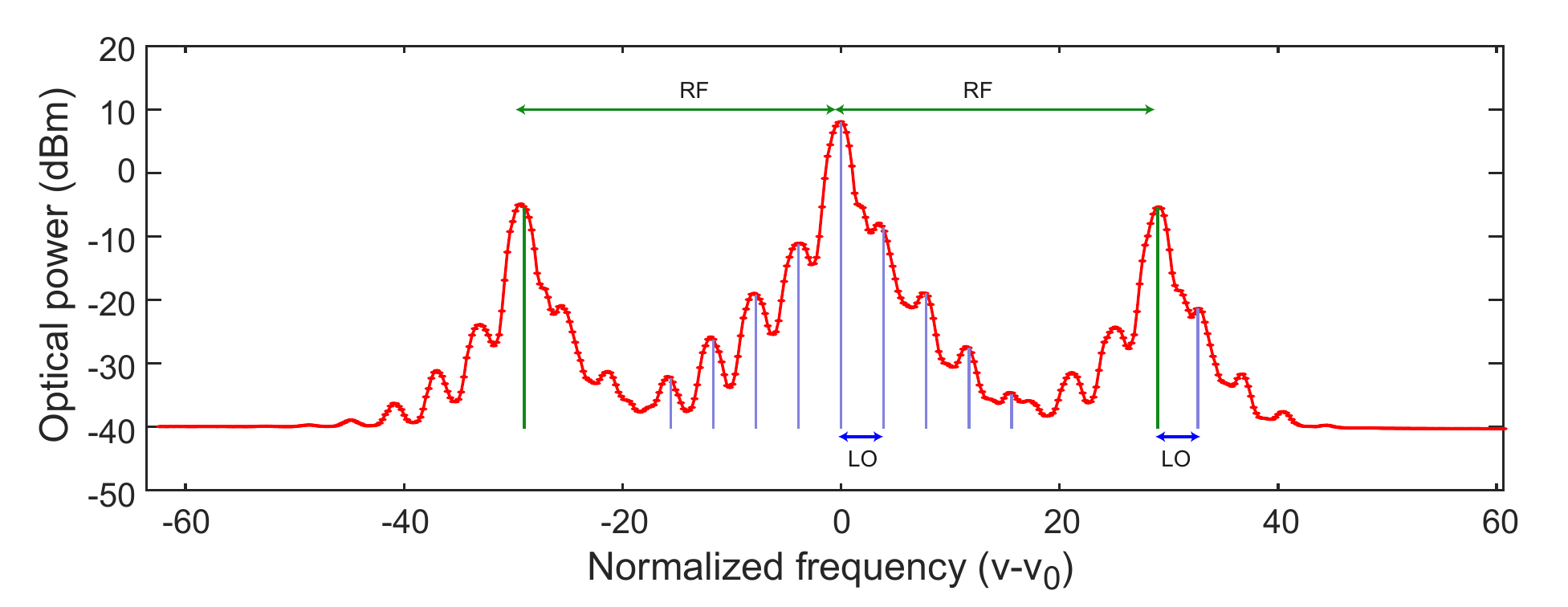}%
%\end{center}
\caption{Recorded optical spectrum \textcolor{Red}{(in red)} when the OMO is sent via the SATCOM OPTIMA test bench. \textcolor{Red}{Vertical lines represent the different harmonics.}}
\label{fig:7extra}
\end{figure}

This section presents the test results of the OMO when mounted in the OPTIMA testbed as described before. Three different configurations are measured to perform a full test of the device as a photonic local oscillator. In all of them, the input signal to PMFC is a 29.0 GHz RF tone with -8 dBm power. An example of the optical power spectrum of the OMO once it is driven through the OPTIMA testbed can be seen in Fig. \ref{fig:7extra}

As reported before, this OMO architecture is highly dependent on \textcolor{Red}{environmental conditions of the room where test bed is installed}, so each time the OMO is switched on, its functional parameters exhibit a slight variation with respect to the previous measurement time. This dependence on the environment \textcolor{Red}{ can strongly affect the excitation conditions}, since the fiber taper that excites the OM cavity can suffer from fluctuations induced by changes in temperature, pressure, or humidity. Notably, improving the coupling mechanisms and getting a packaged device would enormously reduce the environmental dependence, improving the overall performance of the OMO.

\begin{figure}[htbp]
\begin{center}
\includegraphics[width=0.5\textwidth]{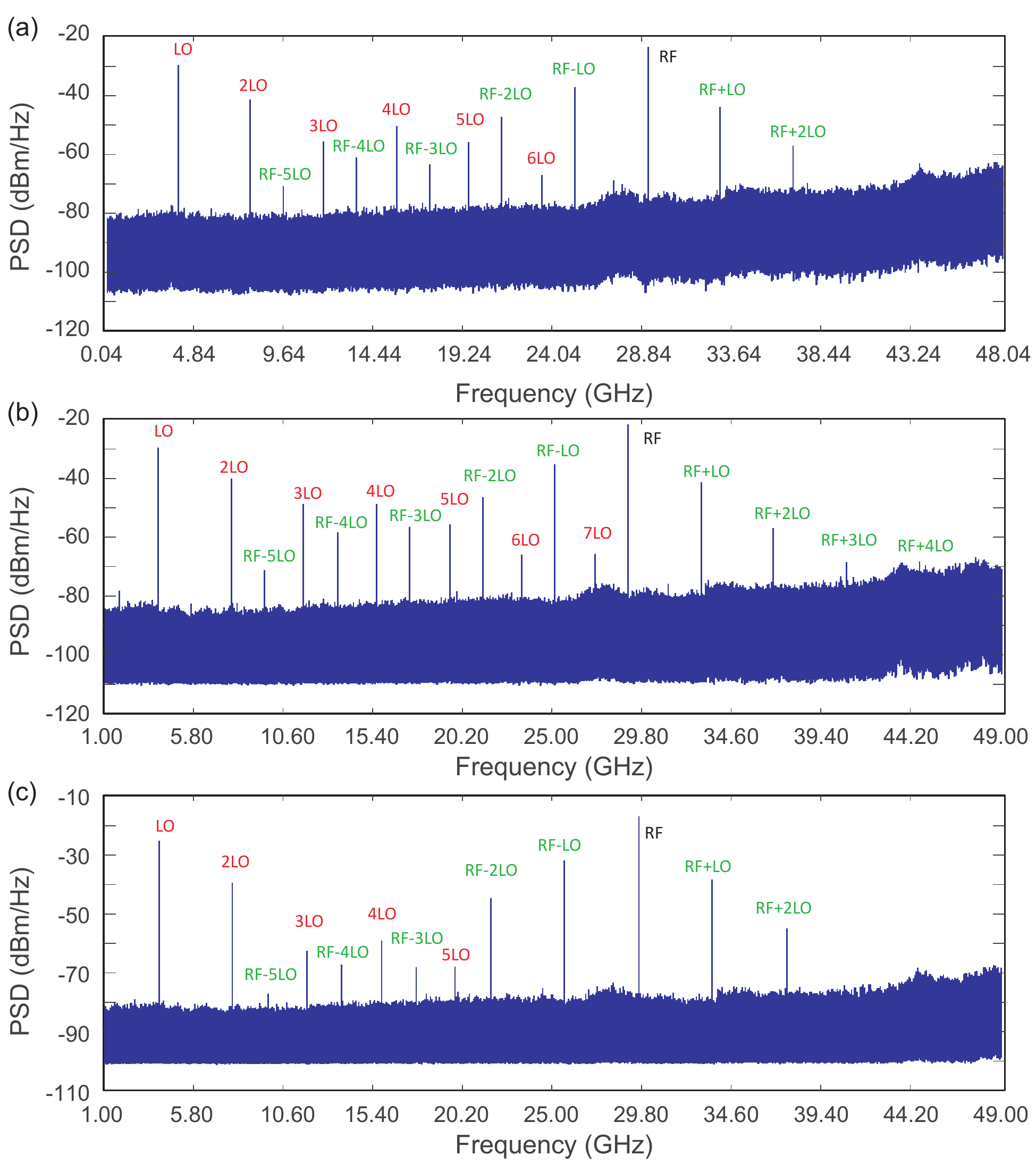}%
\end{center}
\caption{\textcolor{Red}{Output RF spectra for (a) configuration 1 at 1554,59 nm, (b) configuration 2 at 1554,59 nm and (c) configuration 3 at 1558,75 nm.}}
\label{fig:8}
\end{figure}

To provide a reference for the functional tests and the testbed, a set of simulations $–$ using experimental data $-$ was carried out. These simulations consider the performance of each key component at the photonic conversion sub-unit, as well as the known functional parameters already measured on the OMO subsystem (operational wavelength that depends on OMO final stable configuration, LO optical output power that is also dependent on the configuration and the pump power, LO fundamental frequency that depends on the cavity mechanical response, and LO harmonic levels that are dependent on the final configuration). Once the LO optical field is modeled, this signal is amplified by using the physical expressions of EDFA. Then, this LO amplified signal is modulated by using the theoretical expressions of a Mach-Zehnder EOM (insertion loss and half wave voltage ($V_{\pi}$), with a bias voltage as a trade-off between high gain, low noise, and good linearity). Afterwards, two cascaded amplifiers are simulated (the signal is amplified again to boost the signal level and compensate optical losses). Finally, the signal is detected by using a broadband RF photodiode (with responsivity, bandwidth and $S_{21}$ response given by the datasheet; and 50 Ohm impedance on all RF interfaces). All the harmonic and intermodulation products are located along the electrical spectrum to obtain the electrical power of each signal.

%\subsection{Configuration 1}

\textcolor{Red}{It must be noted that the OMO has an optical resonance around 1522 nm with an optical quality factor up to $10^{3}$ \cite{MER20-NN}. However, when the driving power is sufficiently large (as required to reach the phonon lasing regime), the large thermo-optic effect of silicon gives rise to bistability with a ``saw-tooth'' shaped transmission, thus red-shifting the resonance frequency \cite{NAV14-AIPA}. Under this driving condition, the different configurations of the OMO were studied.} In configuration 1 the OMO was stabilized at a wavelength of 1554,59 nm and with an input driving power of 11 mW. This configuration was lost after 3-hour measurement. Then, the OMO was stabilized again at a driving wavelength of 1554,59 nm and with an input power of 11 mW (configuration 2). At other measurement time, in configuration 3, the system was stabilized at a wavelength of 1558,75 nm and an OMO input power equal to 13 mW. The measured output spectrum (for the three configurations) \textcolor{Red}{are shown in Fig. \ref{fig:8}} and summarized in Table \ref{tab:234} (\textcolor{Red}{together} with simulations results of configuration 2).

\begin{figure}[htbp]
\begin{center}
\includegraphics[width=0.5\textwidth]{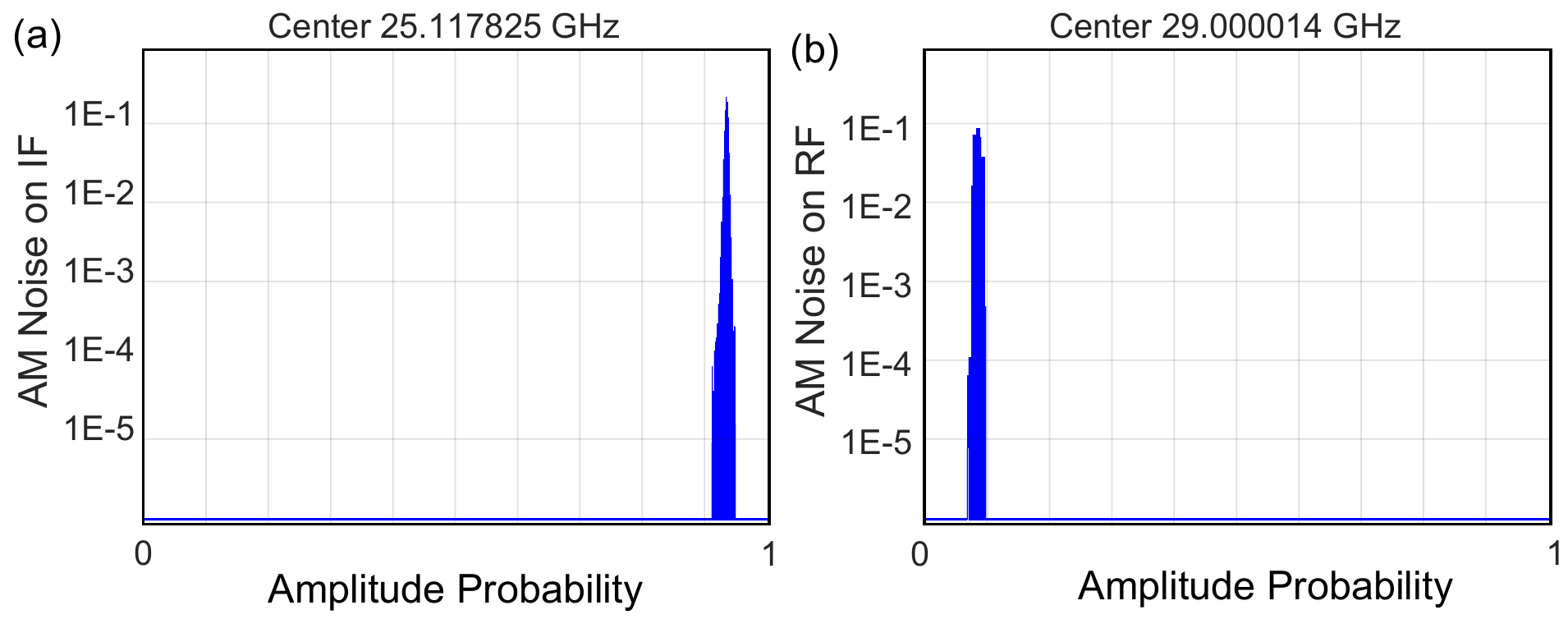}%
\end{center}
\caption{(a) AM noise on IF output signal. (b) AM noise on output RF signal.}
\label{fig:9}
\end{figure}

%\subsection{Configuration 2}

%\subsection{Configuration 3}

\begin{figure}[htbp]
\begin{center}
\includegraphics[width=0.5\textwidth]{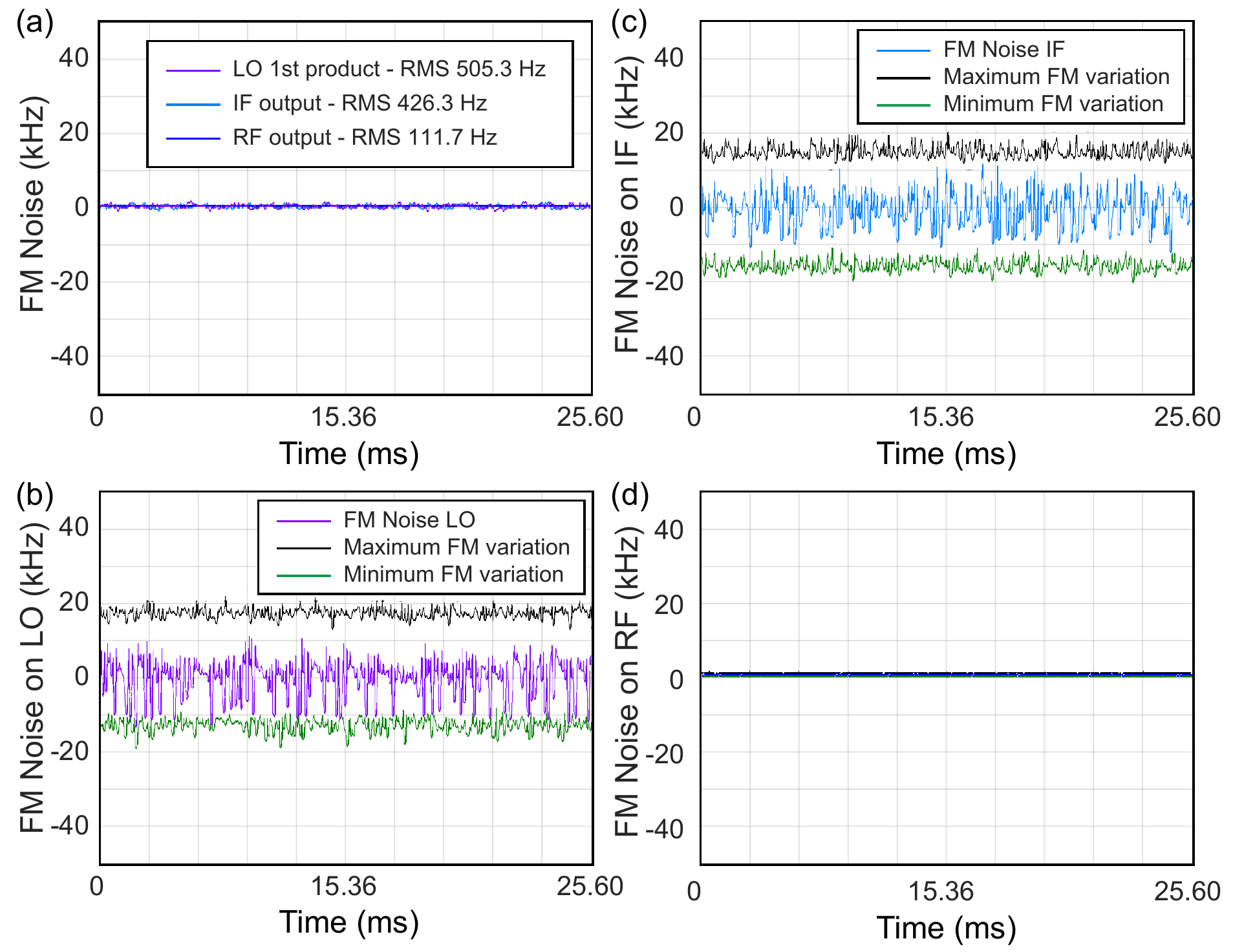}%
\end{center}
\caption{(a) FM noise on LO, IF and RF signal (short term). (b) FM noise on LO first product (long term). (c) FM noise on IF signal (long term). (d) FM noise on RF signal (long term).}
\label{fig:10}
\end{figure}

Besides the spectrum analysis and the intermodulation products, configuration 3 testing was complemented with additional measurements over some output products to assess the amplitude and frequency noise. To this end, the spectrum analyzer was configured as both amplitude modulation (AM) and frequency modulation (FM) analyzer. Figures \ref{fig:9} and \ref{fig:10} show some interesting results of RF, LO and intermediate frequency (IF) output products. \textcolor{Red}{Notice that, in the context of a superheterodyne receiver, the signals are transmitted at the RF frequency and processed at the much-lower IF frequency, being both related by the LO frequency as $f_{LO}=f_{RF}-f_{IF}$.}

\begin{table*}[h]
  \centering
   \caption{Test results of three configurations and simulated values of configuration 2.}
   \label{tab:234}
  \begin{tabular}{@{}cccccccccc@{}}
    \toprule
Product & Freq & Power & Power & Power & Power sim. & Conversion & Conversion & Conversion & Conversion \\ 
        & (GHz) & (dBm) & (dBm) & (dBm) & (dBm)& gain (dB) & gain (dB) & gain (dB) & gain sim. (dB)\\
        &  & Conf. 1 & Conf. 2 & Conf. 3 &   &  Conf. 1 & Conf. 2 & Conf. 3  &   \\
    \midrule
RF &  29 & -25.3 & -22,8 & -18,1& -23.2 & -17.3 &-14,8 & -10,1& -15.2\\
LO &  3,888 & -33,8  & -33,4 &-26,4 & -26,4 &  - & - &-  & - \\
2LO &  7,777 & -45,1 & -36,7  & -39,8& -50,1 &  - &-  &-  &  - \\
3LO &  11,666 & -58,0 & -54,5 & -63,2& -86,3 & -  & - & -& - \\
4LO &  15,554 & -55,6 &-52,1 & -61,4 & -114 & - &- & -& - \\
5LO &  19,443 & -71,7 & -58,4& -69,6& -123 & - &- &- & -\\
RF-LO &  25,111 & -39,3 & -35,4& -32,8& -33,4 & -31,3 & -27,4& -24,8& -25,5\\
RF-2LO &  21,222 & -52,3 & -47 & -46,2& -57,0 & -44,3 &-39,0 & -38,2& -49,0\\
RF-3LO &  17,333 & -65,4 & -56,9&-70,4 & -93,4 & -57,4 &-48,9& -62,4 & -85,4\\
RF-4LO &  13,445 & -70,1 & -59,1& -72,1& -120,5 & -62,1 &-51,1& -64,1 & -112,5\\
RF-5LO &  9,556 & -86,7 & -72,3& -79,4& -119,3 & -78,7 &-64,3 & -71,4& -11,3\\
RF+LO &  32,888 & -45,2 & -42,4& -39,6& -33,4 & -37,2 & -34,4& -31,4& -25,5\\
RF+2LO &  36,777 & -63,3 & -58,4& -58,1& -57,1 & -55,7 &-50,4 &-50,1 & -49,1\\
RF+3LO &  40,666 & -78,7 &-65,6 & -82,6& -93,6 & -70,7 &-70,7 & -74,6& -85,6\\
RF+4LO &  44,554 & -76,4 & -62,9&-77,6 & -113,0 & -68,4 &-68,4 &-69,6 & -105,0\\
    \bottomrule
  \end{tabular}
\end{table*}

To evaluate the AM noise, Fig. \ref{fig:9}(a) depicts the amplitude distribution of the down-converted IF signal. The mean value is -33.09 dBm with a crest factor \textcolor{Red}{(peak power - mean power)} of 1.36 dB, which is an important value since it is an indicator of amplitude instability in the output signal. Next, the same amplitude distribution is shown for the RF output signal in Fig. \ref{fig:9}(b). Here, the mean value is -19.04 dBm with a crest factor of 0.15 dB, which is a more acceptable value. It is an indicator of amplitude stability in the output signal. An early conclusion can be drawn at this point: the main amplitude instability comes from the LO signal, since the mixed IF signal (RF-LO) has more instability than non-mixed output signal (RF).

Another important measurement is the FM noise of the output products. Figure \ref{fig:10}(a) shows the main FM parameters of the LO first harmonic at the output for the three different configurations. This noise was measured during a short time window (short term), typically less than two minutes observation time. The root mean square (RMS) FM noise is 505.3 Hz and the peak-to-peak parameter is 2.8 kHz. This is a considerable value of the frequency noise, which has direct impact in the phase noise. The same value was assessed with a longer observational time (long term, typically 10 minutes). In Fig. \ref{fig:10}(b), two additional traces have been added to bound the maximum and minimum FM variation. Here, the effect of the FM instability is more serious since the RMS FM noise is 4.765 kHz (one order of magnitude higher than before). The peak-to-peak noise was now 24 kHz (also one order of magnitude higher than at short-time measurement).

Also, the IF output signal is FM demodulated and Fig. \ref{fig:10}(a) shows the results for the short term. As with the LO output signal, this noise was measured during a short time window, typically less than two minutes observation time. The RMS FM noise in this case was 426.3Hz and the peak-to-peak parameter is 2.5KHz. These noise levels are similar to the LO noise levels.  The same measurement with a longer time window is carried out, being the result shown in Fig. \ref{fig:10}(c). For a typical 10-minute time-window one has up to 4.5 KHz RMS FM noise and 23KHz peak-to-peak FM noise.
The RF output signal was also monitored, at short term, with the results depicted on Fig. \ref{fig:10}(a). Contrary to LO and IF FM noise levels, RF exhibits less RMS frequency noise (27.39Hz) than its predecessors. The amount of noise is one order of magnitude less than the LO or IF noise. The window period for these measurements was less than two minutes. The same measurement on the RF output level was performed for a longer window period (around 10 minutes), as shown in Fig. \ref{fig:10}(d). The RMS FM noise in this case is 111.7 Hz and a peak-to-peak FM noise of 800 Hz approximately. 

\begin{figure}[htbp]
\begin{center}
\includegraphics[width=0.5\textwidth]{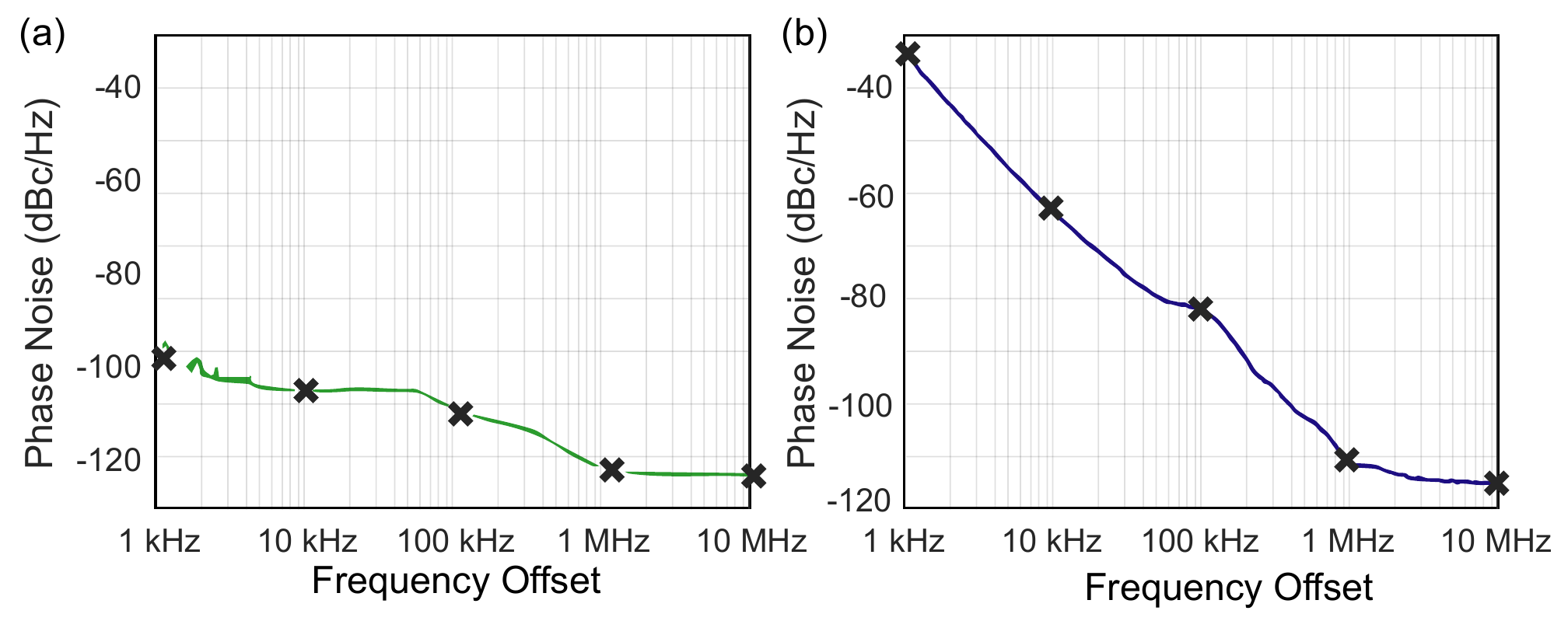}%
\end{center}
\caption{SSB phase noise of the IF signal. (a) OPTIMA PhLO; (b) OMO system.}
\label{fig:11}
\end{figure}

Another important measurement that can be compared with the OPTIMA testbed is the SSB phase noise level on the output IF signal. Figure \ref{fig:11}(a) shows the SSB PN on the OPTIMA IF output signal whilst Fig. \ref{fig:11}(b) shows the same SSB PN but for the IF output signal in the OMO+PMFC testbed. The SSB phase noise is higher in OMO+PMFC test bed than in OPTIMA testbed, especially at lower frequency offsets. The main reason of this is that OMO is not optimized to minimize phase noise. Apart from the dependence of the OM cavity on parameters such as temperature, humidity and mechanical conditions, the optical path is neither properly designed nor integrated to overcome all sources of phase noise, like optical reflections, long fiber pigtails or use of single mode fiber instead of PM fiber. However, this issue may be well solved by improving the design of the cavity as well as its interface with the external system. In this sense, a packaged device is mandatory for practical applications. 

\section{Conclusions}

In summary, we have characterized and discussed a new all-photonic microwave oscillator, essentially performing as an integrated OEO but implemented using an OM cavity built on a silicon chip, in a testbed suitable for validating devices aimed at SATCOM applications. The OM performs as a photonic LO operating approximately at multiples of the mechanical resonance frequency. Looking at the performance as a OMO+PMFC downconverter, first-order down conversion gain (RF-LO) exhibits results comparable to OPTIMA with equivalent parameters \textcolor{Red}{(in this case, -32.8 dBm for both OPTIMA and OMO in configuration 3)}. This is a very positive aspect since it shows that the OMO can \textcolor{Red}{be substitute of} a PhLO, which requires of an external RF source, whilst the OMO generates the LO in a purely optical way. Furthermore, higher-order down-conversion gain (RF-n•LO) shows better results than the OPTIMA system with equivalent parameters \textcolor{Red}{(-75.7 dB and -108 dB for OPTIMA RF-4LO and RF-5LO, respectively)}. This is because the OMO design has a non-linearity factor higher than the EOM-based photonic LO in OPTIMA or similar systems. This is a remarkably good result as well, since higher LO frequencies can be generated to get more frequency conversion bandwidth in a future system architecture. 

One drawback is that the OMO optical sidebands exhibit less power than the typical EOM-based PhLO configuration. This is because the mentioned non-linearity factor in OMO. It gives more energy to the higher LO products, at the expense of less power to the first LO sideband. This has a direct impact in the system conversion gain, being lower in this solution than in the OPTIMA system. About the signal integrity, another improvable two parameters are the amplitude and the frequency stability. It has been seen that the output signal products mixed with LO (n•LO and RF±n•LO) exhibit one order of magnitude more FM noise than non-mixed products (particularly RF output signal). Also, the amplitude noise is higher in the IF signal than in the RF output level. Looking at other relevant parameters, the SSB phase noise is higher in this assessed test board than in the OPTIMA test board.

From a practical perspective, the OMO can significantly reduce the SWaP (size, weight and power) over the conventional approach due to its intrinsically small footprint and the lack of control and biasing electronics. The device is passive and can fit directly in conventional payload configurations (based on Silicon-technology) without the need to implement specific interfaces. In addition, it also provide inherent frequency mixing capabilities as a result of the inherent nonlinear  behavior of the OM cavity \cite{MER21-LPR}. Therefore, with all the advantages, this approach is a very promising candidate to build \textcolor{Red}{ultra-light weight} OEOs, highly appropriate for space applications.

%\appendices

%\section{}

\section*{Acknowledgment}

The authors acknowledge funding from the H2020 Future and Emerging Technologies program (projects PHENOMEN 713450, SIOMO 945915 and \textcolor{Red}{OPTIMA 730149}); the Spanish State Research Agency (PGC2018-094490-BC21 \textcolor{Red}{and ICTS-2017-28-UPV-9}); Generalitat Valenciana (BEST/2020/178, PROMETEO/2019/123, \textcolor{Red}{IDIFEDER/2020/041 and IDIFEDER/2021/061}). %\textcolor{Red}{L.M. also acknowledges the Margarita Salas }

\ifCLASSOPTIONcaptionsoff
  \newpage
\fi

\bibliographystyle{IEEEtran}
%\bibliography{IEEEabrv,biblio}
\bibliography{biblio}

\end{document}